\documentclass[aps,twocolumn,showpacs]{revtex4}
\usepackage{amssymb}
\usepackage{natbib}
\usepackage{amsmath}
\usepackage{amsfonts}
\usepackage{graphicx}
\usepackage{mathrsfs}
\usepackage{dcolumn}
\usepackage{mathbbol}
\usepackage{bm}
\usepackage{color}
\usepackage{epsfig}
\usepackage{units}

\definecolor{rot}{rgb}{0.75,0.05,0.25}
\definecolor{hellgrau}{gray}{0.5}
\definecolor{blau}{rgb}{0,0,0.7}

\def\Tr{\mbox{Tr}}

\begin{document}

\title{Geometric magnetism in open quantum systems}
\author{Michele Campisi, Sergey Denisov, and Peter H\"anggi}
\affiliation{Institute of Physics, University of Augsburg,
  Universit\"atsstr. 1, D-86135 Augsburg, Germany}
\date{\today }

\begin{abstract}
An isolated classical chaotic system, when driven by the slow change of several
parameters, responds with two reaction forces: geometric friction and geometric magnetism.
By using the theory of quantum fluctuation relations we show that this holds true also
for open quantum systems, and provide explicit expressions for those forces in this case.
This extends the concept of Berry curvature to the realm of open quantum systems.
We illustrate our findings by calculating the geometric magnetism of a damped charged quantum harmonic oscillator
transported along a path in physical space in presence of a magnetic field and a thermal environment.
We find that in this  case the geometric magnetism is unaffected by the presence of the heat bath.
\end{abstract}


\maketitle

\section{Introduction}
According to the adiabatic theorem, when a thermally isolated quantum system is subjected to an adiabatic cyclical driving it
returns to the same state where it started from \cite{Messiah62Book}. The only effect of the excursion is that the state acquires a phase. In 1984, Berry clarified that such a phase consists of a
gauge-dependent (hence nonphysical) part and a  gauge-independent (hence physical and measurable) part, which is determined only  by the geometry of the path described by the driving parameters \cite{Berry84PRSLA392}. This is by now customarily called Berry's phase. Soon later Hannay \cite{Hannay85JPA18} and Berry \cite{Berry85JPA18} found the analogous classical
phenomenon: when a classical integrable system undergoes an adiabatic cyclic evolution, the action variable remains constant, but the angle variable experiences an anholonomy effect, that is, it does not return to its original value but accumulates a shift, known as Hannay angle. More recently, Robbins and Berry \cite{Robbins92PRSLA436}
addressed the question of whether there exists a geometric phase also in the case of
 classical chaotic systems.
Their approach was based on the observation that the Berry phase is given by the flux of a two-form
(Berry's curvature) through a surface bounded by the cyclic path in the parameter space.
Therefore, they investigated the classical limit of the quantum two-form and found the expression
\cite{Robbins92PRSLA436}:
\begin{eqnarray}
\bm{\mathcal{B}}^c = \frac{1}{2 \omega(E)} \frac{\partial}{\partial E} \left[ \omega(E) \int_0^{\infty} \mathrm{d}t \langle \bm{\nabla} H_{t=0} \times \bm{\nabla} H_t \rangle_{E} \right]\, ,
\label{eq:clas2form}
\end{eqnarray}
where $\bm{\nabla} H_t$ is the gradient in parameter space of the Hamiltonian evaluated at time $t$ (i.e., the instantaneous negative force exerted by the external driving), $\times$ denotes cross product, $\langle \dots \rangle_{E}$ denotes microcanonical average at energy $E$ and $\omega(E)$ is the corresponding density of states.
Later Jarzynski \cite{Jarzynski95PRL74} showed that the surface integral of the classical chaotic two-form (\ref{eq:clas2form}) measures a shift accumulated along the chaotic trajectory on the constant energy hyper-surface, which generalizes the
concept of Hannay's angle. In a subsequent paper Berry and Robbins \cite{Berry93PRSLA442}
re-derive Eq. (\ref{eq:clas2form}) adopting a statistical mechanical approach. They considered an
initially microcanonically distributed ensemble and focused on the average force with which the system reacts to the external driving, $\bm{F}=- \langle \bm{\nabla} H \rangle_{E}$, up to first order in the driving speed.
They showed that the reaction force contains two terms: a friction-like force and a Lorentz-like force. The latter stems from a magnetic-like field, the so called geometric magnetism,
which is nothing  but the classical chaotic two-form detailed in Eq. (\ref{eq:clas2form}).
The approach developed in  \cite{Berry93PRSLA442} is similar to Kubo's Linear Response Theory \cite{Kubo57aJPSJ12}.
The main differences are: (i) in Kubo's theory the initial state is canonical whereas in Berry and Robbins theory it is microcanonical;
(ii) Kubo's theory gives the response of a driven system to a \emph{weak} perturbation, and gives, accordingly, a linear relationship between the response and the strength of the driving. The theory of Berry and Robbins instead yields the response of a driven system to a \emph{slow} perturbation, and gives, accordingly, a linear relationship between the response and the \emph{speed}
of the driving. We might label Berry and Robbins theory \cite{Berry93PRSLA442} a \emph{Microcanonical Adiabatic Linear Response Theory}.

In the recent years yet further attempts have been devoted to investigate possible generalizations of Berry's phase.
It has been realized that geometric phases can be used as effective and reliable tools for quantum computation \cite{Jones00NAT403,Duan01Science}. Thus, the study of geometric phases in realistic open quantum systems has become of paramount importance. This problem has been typically addressed from a \emph{dynamics} point of view, that is, one researches proper definitions of phases with respect to non-unitary dynamics \cite{Carollo03PRL90,De-Chiara03PRL91,Tong04PRL93,Whitney05PRL94,Buric09PRA80,Bassi06PRA73,Dajka11QIP10,Marzlin04PRL93,Whitney03PRL90,Sinitsyn09JPA42,Ren10PRL104} (which are the relevant ones for open systems), instead of the unitary dynamics originally considered by Berry.

To the best of our knowledge, however, the problem of finding the geometric phase of an open
quantum systems was not previously addressed adopting a statistical mechanical method akin to the
one employed by Berry and Robbins \cite{Berry93PRSLA442}.
Here we pursue this task. Below we develop a \emph{Canonical Adiabatic Linear Response Theory}
for  open quantum systems. As a main result we obtain the general expression of the field of geometric
magnetism of open quantum systems, see Eq. (\ref{eq:main}) below.
The geometric phase is given by its surface integral in analogy with the standard case.
In developing the theory we take full advantage from the theory of quantum work fluctuation relations
\cite{Campisi11RMP83,Esposito09RMP81}, which can be formulated within two complementary
viewpoints: exclusive and inclusive \cite{Jarzynski07CRPHYS8,Campisi11PTRSA369}.
While the former is best suited to derive Kubo's Linear Response Theory
\cite{Bochkov77SPJETP45,Andrieux08PRL100}, the latter, as we will see below, is best suited to
derive the searched Canonical Adiabatic Linear Response Theory.

\section{Adiabatic Linear Response of open quantum systems}
We consider a driven open quantum system in contact with a thermal bath at fixed temperature $T=1/(k_B \beta)$.
Following the established procedure \cite{Zwanzig60JCP33,Feynman63AP24} we
hence close the system by coupling it to a thermal bath.
The total (system+bath) Hamiltonian reads
\begin{equation}
\mathcal{H}(\mathbf{R}_t)= H_B + H_{S B}+H(\mathbf{R}_t) \, ,
\end{equation}
where $H_B$ is the bath Hamiltonian, $H_{S B}$ is the (possibly strong) system/bath coupling, and $H(\mathbf{R}_t)$ is the system Hamiltonian:
\begin{equation}
H(\mathbf{R}_t)= H_0 - \mathbf{R}_t  {\, \cdot\, } \mathbf{Q} \, .
\label{eq:HS}
\end{equation}
Here $t$ is time, $\mathbf{R}_t=(R^1_t, \dots R^N_t)$ denotes a set of time dependent external parameters, (the generalized ``displacements'') $\mathbf{Q}=(Q^1, \dots Q^N)$ is the set of conjugated system  observables (the generalized ``forces''), and ``$ \cdot$'' denotes the scalar product.
We assume that the bath is \emph{ideal}, meaning that it is has infinite heat capacity $C$ and, accordingly, cannot change its temperature $T$ upon injection of finite amounts of energy.
For convenience we introduce the notations
\begin{equation}
\varrho^{\text{eq}}_\mathbf{R}= e^{-\beta \mathcal{H}(\mathbf{R})}/\mathcal{Z}_\mathbf{R},\quad \mathcal{Z}_\mathbf{R}=\text{Tr}\,  e^{-\beta \mathcal{H}(\mathbf{R})} ,
\label{eq:Gibbs}
\end{equation}
to denote the Gibbs equilibrium of the total system at fixed parameters $\mathbf{R}$,
and the corresponding partition function ($\Tr$ is the trace over the total system). We assume that at $t=0$ the total system is in the equilibrium $\varrho^{\text{eq}}_{\mathbf{R}_0}$.
We next consider some system observable $O$, and ask how its expectation value at time $t=\tau$,
\begin{equation}
\langle O_\tau \rangle \doteq \Tr \varrho^{\text{eq}}_{\mathbf{R}_0} O_\tau = \Tr  \varrho_\tau O \, ,
\label{eq:neqexp}
\end{equation}
deviates from its equilibrium expectation value
\begin{equation}
\langle O \rangle_{\mathbf{R}_\tau}^{\text{eq}} \doteq \Tr \varrho^{\text{eq}}_{\mathbf{R}_\tau} O \label{eq:eqexp}\, .
\end{equation}
Here, $O_\tau$ denotes the Heisenberg picture
$
O_t=U_{t 0}^{\dagger}O U_{t 0}
$,
$\varrho_t$ is the total system density matrix at time $t$:
$
\varrho_t=U_{t 0}\varrho_{\mathbf{R}_0}^{eq} U_{t 0}^{\dagger}\,
$,
and $U_{\tau t }$ denotes the quantum time evolution operator from time $t$ to time $\tau$, generated by the total Hamiltonian $\mathcal{H}(\mathbf{R}_t)$.

Using the cyclic property of the trace operator, and the property $U^\dagger_{\tau t}U_{\tau t}=\mathbb{1}$, one can prove the following nonequilibrium identity
\begin{align}
\langle O_\tau e^{-\beta  \mathcal{H}_\tau(\mathbf{R}_\tau)}e^{\beta  \mathcal{H}(\mathbf{R}_0)}
\rangle^{\text{eq}}_{\mathbf{R}_0}= e^{-\beta \Delta F}   \langle O \rangle^{\text{eq}}_{\mathbf{R}_\tau} \, ,
\label{eq:NonEqId}
\end{align}
where $\Delta F=-\beta^{-1}\ln(\mathcal Z_{\mathbf{R}_\tau}/\mathcal Z_{\mathbf{R}_0})$ is the difference of free energy of the total system states $\varrho_{\mathbf{R}_\tau}^{\text{eq}}$ and $\varrho_{\mathbf{R}_0}^{\text{eq}}$.
For $O=\mathbb{1}$ Eq. (\ref{eq:NonEqId})
reproduces the quantum Jarzynski Equality \cite{Talkner07PRE75,Campisi11RMP83}. Eq. (\ref{eq:NonEqId}) may be obtained from the nonequilibrium generating functional of Andrieux and Gaspard  \cite{Andrieux08PRL100} by means of functional differentiation.
Note that the free energy difference
can be written in the following form \cite{Jarzynski97PRL78}:
$
\Delta F = - \int_{\mathbf{R}_0}^{\mathbf{R}_\tau} \mathrm{d}{\mathbf{R}} \cdot \langle \mathbf{Q} \rangle_\mathbf{R}^{\text{eq}}=
 - \int_0^\tau \mathrm{d}t \dot{\mathbf{R}}_t \cdot \langle \mathbf{Q} \rangle_{\mathbf{R}_t}^{\text{eq}}
$.
Using the notations
\begin{align}
W= \mathcal{H}_\tau(\mathbf{R}_\tau)-  \mathcal{H}(\mathbf{R}_0)= - \int_0^\tau \mathrm{d}t \dot{\mathbf{R}}_t \cdot  \mathbf{Q}_t \, , \\
W_{\text{dis}} = W-\Delta F= - \int_0^\tau \mathrm{d}t \dot{\mathbf{R}}_t \cdot  [\mathbf{Q}_t - \langle \mathbf{Q} \rangle_{\mathbf{R}_t}^{\text{eq}}]\ ,
\end{align}
 Eq. (\ref{eq:NonEqId}) reads:
\begin{align}
\langle  O_\tau e^{-\beta  \mathcal{H}_\tau(\mathbf{R}_\tau)}e^{\beta [\mathcal{H}_\tau(\mathbf{R}_\tau) -W_{\text{dis}}]}
\rangle_{\mathbf{R}_0}^{\text{eq}}= \langle O \rangle_{\mathbf{R}_\tau}^{\text{eq}} \, .
\label{eq:NonEqId2}
\end{align}
The operators $W$ and $W_{\text{dis}}$ do not correspond to any quantum-mechanical observable \cite{Talkner07PRE75,Campisi11RMP83}, but approach, -- in the classical limit --, the exclusive work, $w$, and dissipated work, $w-\Delta F$, respectively \cite{Campisi11PTRSA369}.
Under our assumptions  that the bath has infinite heat capacity,
the nonequilibrium expectation (\ref{eq:neqexp}) of
 $W_{\text{dis}}$ vanishes in the adiabatic limit  (for a discussion of the scaling of $W_{\text{dis}}$ with the bath size in a classical set-up, see  \cite{Vaikuntanathan09EPL87}).
Since the expectation of $W_{\text{dis}}$ is given by the Kullback-Leibler relative entropy between $\varrho_\tau$ and $\varrho_{\mathbf{R}_\tau}^{\text{eq}}$ \cite{Schloegl66ZP191,Bochkov81aPHYSA106,Kawai07PRL98,Deffner10PRL105}, this also means that
in the adiabatic limit $\varrho_\tau\rightarrow \varrho_{\mathbf{R}_\tau}^{\text{eq}}$.
As the driving speed increases, the actual state $\varrho_\tau$ lags more and more behind the ``reference'' equilibrium state $\varrho_{\mathbf{R}_\tau}^{\text{eq}}$\cite{Vaikuntanathan09EPL87}.

In the following we consider slow (quasi-adiabatic) driving, and accordingly expand Eq. (\ref{eq:NonEqId2}) to first order in $W_{\text{dis}}$ (here slow means that the characteristic time of variation of the driving is small compared to the characteristic time of relaxation to the Gibbs equilibrium).
Following the method outlined in Ref. \cite{Andrieux08PRL100}, we use the operator expansion:
\begin{equation}
 e^{\beta A} e^{-\beta (A-\epsilon)} =  1+\int_0^\beta \mathrm{d}u \, e^{uA} \, \epsilon \,  e^{-uA} +O(\epsilon^2) \, .
\end{equation}
Setting $A=-  \mathcal{H}_{\tau}(\mathbf{R}_\tau)$ and $\epsilon= -W_{\text{dis}}$, we arrive, to first order in $W_{\text{dis}}$, at the result
\begin{align}
\langle & \Delta  O_\tau  \rangle \doteq \langle  O_\tau \rangle  - \langle  O \rangle_{\mathbf{R}_\tau}^{\text{eq}}   \label{eq:NonEqIdLin3}\\
 &=-  \int_0^\tau \mathrm{d}t \int_0^\beta \mathrm{d} u  \langle  O_\tau  e^{-u \mathcal{H}_\tau(\mathbf{R}_\tau)} \Delta \mathbf{Q}_t e^{u \mathcal{H}_\tau(\mathbf{R}_\tau)} \cdot \dot{\mathbf{R}}_t \rangle^{\text{eq}}_{\mathbf{R}_0} \, .\nonumber
\end{align}
Using the cyclic property of the trace operator and unitarity $U^\dagger_{\tau t}U_{\tau t}=\mathbb{1}$,
we rewrite the integrand in Eq. (\ref{eq:NonEqIdLin3}) as:
\begin{align}
\Tr  \varrho_\tau  O U_{\tau 0} e^{-u \mathcal{H}(\mathbf{R}_\tau)} U_{\tau t}
\Delta  \mathbf{Q} U_{\tau t}^\dagger
e^{u \mathcal{H}(\mathbf{R}_\tau)} U_{\tau 0}^\dagger
\cdot \dot{\mathbf{R}}_t \, .
\end{align}
Since this expression is already of first order in $W_{\text{dis}}$ we can
replace the exact density matrix $\varrho_\tau$ with the approximate equilibrium
density matrix $\varrho^{\text{eq}}_{\mathbf{R}_\tau}$.
The next, crucial assumption is that the correlation function in Eq. (\ref{eq:NonEqIdLin3}) decays quickly compared to the time scale of variation of $\mathbf{R}_t$, which, in fact was assumed to be very large.
Under this assumption, one can approximate the exact time evolution operator $U_{\tau t}$ with the time evolution at frozen $\mathbf{R}=\mathbf{R}_\tau$:
$
U_{\tau t} \simeq e^{-i  \mathcal{H}(\mathbf{R}_\tau)(t-\tau)/\hbar}\, ,
$
replace $\mathbf{R}_t$ by $\mathbf{R}_\tau$, to arrive at:
\begin{align}
\langle \Delta  O_{\tau} \rangle &=-
\sum_i \int_{0}^\tau \mathrm{d}t
{\Phi}_{i,O}^{\mathbf{R}_\tau}(t-\tau) \,
 \dot{R}_\tau^i
\label{eq:NonEqIdLin5} \\
{\Phi}_{i,O}^{\mathbf{R}_\tau}(t) & \doteq
\int_0^\beta \mathrm{d} u
\langle \Delta O_{-i\hbar u} \Delta Q^i_{t} \rangle_{\mathbf{R}_\tau}^{\text{eq}}\, .
\label{eq:corr}
\end{align}
Note that ${\Phi}_{i,O}^{\mathbf{R}_\tau}(t)$ is the quantum equilibrium correlation function between $O$ and $\Delta Q^i$ (i.e., the relaxation function \cite{Kubo57aJPSJ12})
calculated with respect to the
equilibrium state and propagator at fixed $\mathbf{R}=\mathbf{R}_\tau$.

\section{Geometric friction and geometric magnetism}
The theory applies regardless of the number $N$ of driving parameters.
Geometric magnetism only appears in the case where there are at least $N\geq 2$ driving parameters.

Choosing $O$ as the $i$-th component of the force, $Q^i$, Eq. (\ref{eq:NonEqIdLin3}) becomes, using vector notation:
\begin{align}
\langle \Delta  \mathbf{Q}_\tau \rangle &= -\mathbf{K}(\mathbf{R}_\tau) \, \dot{\mathbf{R}}_\tau  \, ,
\label{eq:NonEqIdLin4}
\end{align}
where $\mathbf{K}(\mathbf{R}_\tau)$ is the $N\times N$ conductance matrix whose elements are the integrated force-force equilibrium correlation functions:
\begin{align}
&K^{jk}(\mathbf{R}_\tau)= \int_{0}^\tau \mathrm{d}t \int_0^\beta \mathrm{d} u  \langle \Delta Q^j_{-i\hbar u} \Delta Q^k_{t-\tau} \rangle_{\mathbf{R}_\tau}^{\text{eq}}\\
&= \int_{0}^\tau \mathrm{d}t \left(\int_0^\beta \mathrm{d} u  \langle Q^j_{-i\hbar u}  Q^k_{t-\tau} \rangle_{\mathbf{R}_\tau}^{\text{eq}}
-\langle Q^j \rangle_{\mathbf{R}_\tau}^{\text{eq}}\langle Q^k\rangle_{\mathbf{R}_\tau}^{\text{eq}} \right)\, .
\label{eq:NonEqIdLin6}
\end{align}
Note that the r.h.s of  Eqs. (\ref{eq:NonEqIdLin5}, \ref{eq:NonEqIdLin4}) are geometric, i.e., they depend on time only through the time dependent parameters $\mathbf{R}_\tau$.
Following Berry and Robbins \cite{Berry93PRSLA442} we assume for simplicity a parameter space of dimension $N=3$ and rewrite Eq. (\ref{eq:NonEqIdLin4}) in vector notation as:
\begin{align}
\langle \Delta  \mathbf{Q}_\tau \rangle
&= -\mathbf{K}^\text{S}(\mathbf{R}_\tau) \dot{\mathbf{R}}_\tau -{\bm{\mathcal{B}}}(\mathbf{R}_\tau) \times \mathcal{\dot{\mathbf{R}}_\tau}\, .
\label{eq:NonEqIdLin7}
\end{align}
The first term, stemming from the symmetric part $\mathbf{K}^{\text{S}}$ of the conductance matrix $\mathbf K$, is \emph{geometric friction}, and the second term, stemming from the antisymmetric part $\mathbf{K}^{\text{A}}$ of the conductance matrix is \emph{geometric magnetism}. The field of geometric magnetism, $\bm{\mathcal{B}}$, has components $\mathcal{B}_i= \nicefrac{-1}{2} \sum_{jk}\varepsilon_{ijk}K^{jk}$ [$\varepsilon_{ijk}$ is the Levi-Civita tensor], and reads, in vector notation:
\begin{equation}
{\bm{\mathcal{B}}}(\mathbf{R}_\tau) = \nicefrac{-1}{2} \int_{0}^\tau \mathrm{d}t  \int_0^\beta \mathrm{d} u  \langle \mathbf{Q}_{-i \hbar u} \times  \mathbf{Q}_{t-\tau}\rangle_{\mathbf{R}_\tau}^{\text{eq}}\, .
\end{equation}

The theory may be generalized to account for general \emph{nonlinear} driving, i.e., to system Hamiltonians $H(\mathbf{R}_t)$ not necessarily of the form in Eq. (\ref{eq:HS}). In the general case the operator $W$ reads \cite{Jarzynski07CRPHYS8}:
$
W = \int_0^\tau \mathrm{d}t \dot{\mathbf{R}}_t \cdot \bm{\nabla} H_{t}(\mathbf{R}_t)
$. Accordingly, the general theory is obtained by replacing everywhere $\mathbf{Q}$ with $\bm{\nabla} H$.
A main result, therefore, is that the field of geometric magnetism emerges as:
\begin{equation}
{\bm{\mathcal{B}}}(\mathbf{R}) = \nicefrac{1}{2} \int_0^{\infty} \mathrm{d}t  \int_0^\beta \mathrm{d} u  \langle \bm{\nabla}H_{-i \hbar u} \times \bm{\nabla}H_{t}\rangle_{\mathbf{R}}^{\text{eq}}\, .
\label{eq:main}
\end{equation}
where for simplicity we have changed the integration domain from $[0,\tau]$ to $(-\infty,0]$,
we have used the fact that the antisymmetric part of the relaxation function is odd under
$t\rightarrow -t$ \cite{Kubo57aJPSJ12} to express the result as an integral from $0$ to $\infty$,
and we have dropped the time label $\tau=0$ in ${\bm R}$.
Eq. (\ref{eq:main}) is the \emph{open quantum system} version of Berry and Robbins expression (\ref{eq:clas2form})
for the geometric magnetism of an isolated classical chaotic system.

It is worthwhile to re-express Eq. (\ref{eq:main}) in terms of the symmetrized force autocorrelation
function
\begin{equation}
\Psi^{\bm R}_{jk}(t)= \nicefrac{1}{2}\langle \{ \partial_k H , \partial_j H_t\} \rangle_{\mathbf{R}}^{\text{eq}}\, ,
\label{eq:Psi}
\end{equation}
where $\{\cdot,\cdot\}$ denotes quantum anti-commutator  and $\partial_j = \partial/\partial R_j$.
To this end we rewrite it in tensor notation as
\begin{equation}
\mathcal{B}_i({\bm R})= (\nicefrac{-1}{2})\sum_{jk} \varepsilon_{ijk} \int_{0}^{\infty} \mathrm{d}t
\Phi^{{\bm R}}_{jk}(t)\, ,
\label{eq:main-tensor0}
\end{equation}
where $\Phi^{{\bm R}}_{jk}(t)$ is the relaxation function between $\partial_j H$ and $\partial_k H$,
see Eq. (\ref{eq:corr}). According to the fluctuation-dissipation theorem \cite{Kubo57aJPSJ12}
\begin{align}
\Phi^{\bm R}_{jk} (t) &= \int_{-\infty}^{\infty} \mathrm{d} t'\Gamma(t-t') \Psi^{\bm R}_{jk}(t')\\
\Gamma(t) &= \frac{2}{\hbar \pi} \ln \left[\coth\left(\frac{\pi |t|}{2\beta \hbar} \right)\right]\, ,
\end{align}
hence
\begin{align}
\mathcal{B}_i({\bm R})= \nicefrac{-1}{2} \sum_{jk} \varepsilon_{ijk}  \int_0^{\infty} \mathrm{d}t
\int_{-\infty}^{+\infty}\mathrm{d}t'  \Gamma(t-t') \Psi^{\bm R}_{jk}(t')\, ,
\label{eq:main-tensor}
\end{align}
or, in vector notation:
\begin{align}
\bm{\mathcal{B}}({\bm R}) =\nicefrac{1}{4} &   \int_0^{\infty} \mathrm{d}t
\int_{-\infty}^{+\infty}\mathrm{d}t' \Gamma(t-t')\nonumber \\
&\langle {\bm \nabla H}\times  {\bm \nabla H}_{t'}  -  {\bm \nabla H}_{t'}\times  {\bm \nabla H} \rangle_{\mathbf{R}}^{\text{eq}}\, .
\label{eq:main2}
\end{align}

Eq. (\ref{eq:main-tensor0}) can be re-written in a remarkably simple form:
\begin{equation}
\mathcal{B}_i({\bm R})= (\nicefrac{-1}{2})\sum_{jk} \varepsilon_{ijk} \widetilde \Phi^{{\bm R}}_{jk}(0)\, ,
\label{eq:main-tensor-laplace}
\end{equation}
where the symbol $\widetilde \Phi^{{\bm R}}_{jk}(s)= \int_0^\infty \mathrm{d}t \, e^{-st}\Phi^{{\bm R}}_{jk}(t)$ denotes the Laplace transform of $\Phi^{{\bm R}}_{jk}(t)$.

\subsection*{Classical limit}
Eq. (\ref{eq:NonEqIdLin4}) holds true also classically. The derivation can be repeated following the quantum derivation given above, allowing
observables to commute. As a result, the quantum thermal correlation functions have to be replaced by the classical expressions  \cite{Kubo57aJPSJ12}, so that the classical geometric magnetism reads:
\begin{equation}
{\bm{\mathcal{B}}}^{\rm{cl}}(\mathbf{R}) = \nicefrac{\beta}{2}  \int_0^{\infty} \mathrm{d}t  \langle \bm{\nabla}H \times \bm{\nabla}H_{t}\rangle_{\mathbf{R}}^{\text{eq}} \, .
\label{eq:geomagnClass}
\end{equation}
This result may also be obtained by taking the limit $\hbar\rightarrow 0$ of Eq. (\ref{eq:main}).
Alternatively one can take the limit $\hbar\rightarrow 0$ of Eq. (\ref{eq:main2}). In this limit
$\Gamma(t)\rightarrow \beta \delta(t)$ \cite{Kubo57aJPSJ12}, where $\delta$ denotes Dirac's delta function,
and observables commute
$   {\bm \nabla H}_{t'}\times  {\bm \nabla H}\rightarrow  - {\bm \nabla H}\times  {\bm \nabla H}_{t'}$.

\section{Open System Dynamics}
\label{open}
The geometric magnetism (and the geometric friction too)
may be recast in the more familiar language of \emph{dissipative open system dynamics} 
\cite{Grabert88PREP168,Hanggi05CHAOS15},
in terms of the system reduced density matrix
\begin{align}
\rho_\tau^S & = \Tr_B \rho_\tau\, ,
\end{align}
where $\Tr_B$ denotes the trace over the bath Hilbert space.
The linear response of the force, which defines the conductance matrix ${\bm K}$ (hence the geometric friction and the geometric magnetism), may be written as:
\begin{align}
\langle \Delta  {\bm Q}_\tau \rangle &=  \Tr \rho_\tau   {\bm Q} - \Tr \rho_{{\bm R}_\tau}^{\text{eq}}   {\bm Q} \nonumber \\
&= \Tr_S \rho_\tau^S   {\bm Q} -\langle   {\bm Q} \rangle_{{\bm R}_\tau}^{\text{eq},S} \, ,
\end{align}
where $\Tr_S$ denotes the trace over the system-$S$ Hilbert space and $\langle \cdot \rangle_{{\bm R}}^{\text{eq},S}$ denotes expectation over the equilibrium reduced density matrix  \cite{Campisi09PRL102}
\begin{align}
 \rho_{\bm R}^{\text{eq},S} &= \Tr_B e^{-\beta \mathcal{H}({\bm R})}/\mathcal{Z}_{\bm R}=
e^{-\beta H^*({\bm R})}/Z_{\bm R} \, ,
\label{eq:reduced2}
\end{align}
where $H^*({\bm R})$ is the Hamiltonian of mean force and $Z_{\bm R} = \Tr_S  e^{-\beta H^*({\bm R})}$ is the partition function of an open quantum system \cite{Feynman63AP24,Grabert88PREP168,Hanggi05CHAOS15,Campisi09PRL102}.
In the case of weak coupling the Hamiltonian of mean force reduces to the system Hamiltonian $H_S$.

The element $K^{jk}$ of the conductance matrix may be experimentally/numerically
obtained by driving the system with a small constant
velocity in the $\hat{\bm j}$ direction, $V_j \hat{\bm j}$
and measuring/computing the reaction force in the $k$ direction:
\begin{equation}
K^{jk}(\bm{R}_\tau) =- \left[\Tr_S \rho_\tau^{S,j}  Q_k -\langle   Q_k \rangle_{{\bm R}_\tau}^{\text{eq},S}\right]/{V_j} \, ,
\end{equation}
where we have introduced the notation $\rho_\tau^{S,j} $ to denote the reduced density matrix
resulting from the perturbation $V_j \hat{\bm j}$.
Accordingly, the geometric magnetism may be expressed in terms of the reduced density matrix
\begin{equation}
\mathcal{B}_i({\bm R}_\tau) = \nicefrac{1}{2}\sum_{jk}\varepsilon_{ijk}  \left[\Tr_S \rho_\tau^{S,j}  Q_k -\langle   Q_k \rangle_{{\bm R}_\tau}^{\text{eq},S}\right]/{V_j}\, .
\label{eq:GeoMagRed}
\end{equation}

As illustrated above the geometric magnetism may be accessed
also by calculating the equilibrium force autocorrelation function.
Although quantum correlation functions can not in general be expressed as expectations
over the reduced density matrix, they are ``open quantum systems'' objects that depend
explicitly on bath properties, notably the bath spectral density, see Eq. (\ref{eq:tildePhi-DHO})
below. 
In particular, it should be noted that exact open quantum system dynamics generally is (i)
neither linear \cite{Romero04PRA69}  (ii) nor can it be described by trace preserving completely positive maps 
\cite{Pechukas94PRL73,Pechukas95PRL75}.
 Attempts to resort to approximations (e.g. Markov, and rotating wave) to express the correlation functions in terms
of Markovian dynamics for the system observables, may lead to results which contradict basic principles such as
the fluctuation dissipation theorem \cite{Talkner86AP167}, and therefore to non-negligible errors
in the evaluation of geometric friction and magnetism. See Ref. \cite{Larson12PRL108}
for a recent example of the drastic effects that even good approximations may have on the calculation
of geometric phases.
Therefore a very special care must be paid when employing such approximations in this context.

\section{Illustration: the damped charged harmonic oscillator in a magnetic field}
As an illustration of the theory we consider a quantum damped charged harmonic oscillator
of mass $m$ and charge $q$ transported along a path ${\bm R}_t$ in presence of a constant magnetic field $\bm{B}$.
Adopting the Caldeira-Leggett model of quantum Brownian motion \cite{Caldeira83AP149},
the system, bath and coupling Hamiltonian read:
\begin{align}
H(\bm{R}_t) &= ({\bm p -q \bm A})^2/(2m)+ m\omega^2 {\bm x}^2/2 -  m\omega^2 {\bm x}\cdot {\bm R}_t
 \nonumber \\
 H_B&= \sum_{n=1}^N [{\bm P}_n^2/m_n +m_n\omega_n^2{\bm \xi}_n^2 ]/2\nonumber \\
H_{SB}&= - {\bm x} \cdot \sum_{n=1}^N c_n {\bm \xi_n} + {\bm x}^2\sum_{n=1}^N c_n / (2m_n\omega_n^2)\, .
\label{eq:HO}
\end{align}
Here ${\bm x}, {\bm p}, \omega$ denote the harmonic oscillator
position, momentum and frequency, respectively.
${\bm \xi_n}, {\bm P_n}, m_n, \omega_n$ denote the $n$th bath's oscillator
position, momentum, mass and frequency, respectively. The symbol $c_n$ denotes
the linear coupling constant between the harmonic oscillator and the $n$th bath's oscillator.
The symbol ${\bm A}$ denotes the vector potential. Note that according to Eq. (\ref{eq:HS}) ${\bm Q} =m \omega^2{\bm x}$.
Assuming an initial global Gibbs distribution and adopting the Feynmann-Vernon
path integral approach \cite{Grabert88PREP168} one arrives, after integrating out the bath's degrees of freedom, at the following generalized quantum Langevin equation for the charged oscillator's position \cite{Li90PRA42}
\begin{align}
m \ddot{\bm x}_t+ \int_{-\infty}^t \mathrm{d}t'\eta(t-t')\dot{\bm x}_{t'}-q\dot{\bm x}_t\times {\bm B}+m\omega^2{\bm x}_t={\bm F}_t+{\bm f}_t\, ,
\label{eq:GQLE}
\end{align}
where $\eta(t)$ is the friction kernel, ${\bm f_t}= m\omega^2 {\bm R}_t$ is the externally applied force,
and ${\bm F}_t$ is the stochastic force.
Without loss of generality  we shall assume that $\bm{B}$ points in the $z$ direction, $\bm{B}=B\hat{\bm z}$.
Since the motion in the $z$ direction is decoupled from the motion in the $x$ and $y$
directions, the $xz$ and $yz$ relaxation functions vanish, implying that the geometric
magnetism is also directed in the $\hat{\bm z}$ direction. Further, due to spatial homogeneity,
geometric magnetism does not depend explicitly on the position $\bm{R}_t$.
That is, $\bm{\mathcal{B}}= \mathcal{B} \hat{\bm z}$. From
the compact expression (\ref{eq:main-tensor-laplace}) the strength of the geometric magnetism reads
\begin{align}
\mathcal{B}
&= -\nicefrac{1}{2} [\widetilde \Phi_{xy}(0)- \widetilde \Phi_{xy}(0)] = -\widetilde \Phi^{a}_{xy}(0)
\label{eq:B=Laplace}
\end{align}
where we have introduced the notation $ \Phi_{xy}^{a}$ for the antisymmetric component of $\Phi_{xy}$
and $\widetilde \Phi_{xy}^a(s)$ for its Laplace transform.
Following \cite{Li90PRA42} the Laplace transform of the antisymmetric part of the response function
 reads
\begin{equation}
\tilde \alpha_{xy}^a(s) =  \frac{(m\omega^2)^2\,  qB s}{[m\omega^2+ms^2+s\tilde \eta(s)]^2+q^2 B^2s^2}\, ,
\label{eq:tildeAlphaXY}
\end{equation}
where $[\cdot,\cdot]$ denotes quantum commutator, and
$\tilde \eta(s)$ is the Laplace transform of the bath friction kernel.
Its form depends on the bath spectral density.
For an ohmic bath $\tilde \eta(s)$ is constant.
As compared to Eq. (2.15) of Ref. \cite{Li90PRA42}, we have in our Eq. (\ref{eq:tildeAlphaXY})
an extra factor $(m\omega^2)^2$ stemming from our definition of $\alpha_{xy}^a$ in terms of
$Q_x=m\omega^2 x, Q_y=m\omega^2 y$, rather than $x,y$.

Since $\Phi_{xy}=\int_t^\infty \mathrm{d}t' \alpha_{xy}(t')$ \cite{Kubo57aJPSJ12} we have
\begin{align}
\tilde \Phi_{xy}^a(s) &= \frac{\widetilde \alpha_{xy}^a(0)}{s} - \frac{(m\omega^2)^2\,  qB }{[m\omega^2+ms^2+s\tilde \eta(s)]^2+q^2 B^2s^2} \nonumber \\
&=- \frac{(m\omega^2)^2\,  qB }{[m\omega^2+ms^2+s\tilde \eta(s)]^2+q^2B^2s^2}\, ,
\label{eq:tildePhi-DHO}
\end{align}
where $\alpha_{xy}^a(0)=0$ due to the fact that at equal times $Q_x$ and $Q_y$ commute.
Regardless of the bath spectral density, the friction kernel $\eta(t)$ vanishes at large times,
hence, according to the final value theorem  $\lim_{s\rightarrow 0} s\tilde \eta (s)=0$.
Using Eq. (\ref{eq:B=Laplace}) one finally obtains the result
\begin{align}
\mathcal{B}=q B\, ,
\label{eq:B=qB}
\end{align}
which evidently holds both classically and quantum-mechanically.
Apart from the charge $q$, geometric magnetism is nothing but the physical magnetic field in this case. The factor $q$ stems from the fact that the geometric Lorentz force in Eq. (\ref{eq:NonEqIdLin7}) reads $-\bm{\mathcal{B}}\times\dot{\bm{R}}$, whereas the Lorentz force reads $-qB \times v$ [where $v=(\dot x, \dot y, \dot z)$].
This very same result was found also in Refs.
\cite{Robbins94JPA27,Exner00PLA276} for the case of an
isolated  classical or quantum harmonic oscillator.
Our result (\ref{eq:B=qB}) conveys the non-trivial knowledge that
this continues to hold also for an open classical or quantum harmonic oscillator.
That is, the presence of a bath does not destroy the geometric magnetism,
in fact it does not minimally alter it in this case.
Analogous calculations involving the symmetric part of the relaxation function
lead to the result that the geometric friction is given by the time integral of the
friction kernel $\int_0^\infty  \mathrm{d}t \eta (t)$.

It is noteworthy that the case of geometric magnetism is distinct from the case of standard equilibrium diamagnetism, which is
absent in the  classical limit of open systems and reveals itself at the quantum level only, see Bohr-van Leeuwen  theorem \cite{FeynmannBook,Pradhan10EPL89}.

\section{Concluding Remarks}

We have derived a general expression for the field of geometric magnetism in open quantum systems, Eq.  (\ref{eq:main}),
possibly coupled strongly to the environment. This generalizes the expression (\ref{eq:clas2form}) of Berry and Robbins \cite{Berry93PRSLA442,Robbins92PRSLA436} which refers to closed chaotic classical systems.
It is worth noticing that, contrary to the case studied by Berry and Robbins, here no assumption of chaotic dynamics of the driven system $H(\bm{R}_t)$, which may well be integrable, is made. It is the presence of the thermal bath $H_B$ and the coupling to it, $H_{BS}$, that provide the necessary degree of chaos for the development of a response theory \`a la Kubo.
It is however important to remark the differences between the presently developed theory and that of Kubo.
This is best seen by confronting Eq. (\ref{eq:NonEqIdLin5}) with Kubo's formula
\begin{equation}
\langle O_\tau \rangle - \langle O \rangle_{\bm{R}_0}^\text{eq} =
 \int_0^\tau \mathrm{d}t \int_0^\beta \mathrm{d} u
\langle \Delta O_{-i\hbar u} \Delta \dot{\mathbf{Q}}_{t-\tau} \rangle^{\rm{eq}}_{\bm{R}_0} \cdot \mathbf{R}_t\, .
\label{eq:kubo}
\end{equation}
Note the prominent difference that Kubo's formula (\ref{eq:kubo}) gives an expression (linear in $\mathbf{R}$) for the difference between the nonequilibrium expectation of  $O$ at time $\tau$, and its equilibrium expectation at time $0$, whereas the present formula (\ref{eq:NonEqIdLin5}) gives an expression (linear in $\dot{\mathbf{R}}$) for the difference between the  nonequilibrium expectation of  $O$ at time $\tau$, and its equilibrium expectation \emph{at the same time} $\tau$. Thus in Kubo's theory the small parameter is the \emph{strength} of the driving whereas in our theory the small parameter is the \emph{speed}. Both formulae (\ref{eq:NonEqIdLin5}, \ref{eq:kubo}) yield the response in terms of equilibrium correlation functions. While Kubo's formula involves the correlation between $O$ and $\dot Q$ (the response function), our formula involves the correlation between $O$ and $Q$ (the relaxation function). Note that Kubo's formula (\ref{eq:kubo}) follows from an exact  fluctuation relation
\begin{equation}
\langle O_\tau e^{-\beta  H_{0,\tau}}e^{\beta  {H}_0}
\rangle^{\text{eq}}_{\mathbf{R}_0}=   \langle O \rangle^{\text{eq}}_{\mathbf{R}_0}\, ,
\label{eq:BKintegralFT}
\end{equation}
that looks very similar to our starting Eq. (\ref{eq:NonEqId}) \cite{Bochkov77SPJETP45,Andrieux08PRL100}.
The differences are that (i) the r.h.s. is evaluated at $\bm{R}_0$ in Eq. (\ref{eq:BKintegralFT}), while it is calculated at $\bm{R}_\tau$ in  Eq. (\ref{eq:NonEqId}), (ii) Eq. (\ref{eq:BKintegralFT}) does not involve the free energy difference $\Delta F$, which instead appears in Eq. (\ref{eq:NonEqId}), (iii)
in Eq. (\ref{eq:BKintegralFT}) the unperturbed system Hamiltonian $H_0$
appears instead of the total Hamiltonian $\mathcal H(\bm{R}_t)$ appearing in Eq. (\ref{eq:NonEqId}).
These complementary expressions (\ref{eq:NonEqId}) and (\ref{eq:BKintegralFT}) are customarily referred to as ``inclusive viewpoint'' and ``exclusive viewpoint'' fluctuation relations, respectively.
Interested readers can find accounts of the importance of these viewpoints in the theory of nonequilibrium fluctuations in Refs. \cite{Jarzynski07CRPHYS8,Campisi11RMP83,Campisi11PTRSA369}.
Just like Eq. (\ref{eq:BKintegralFT}) allows one to obtain Kubo's formula (\ref{eq:kubo}) and the whole hierarchy of higher order nonlinear responses, so does Eq. (\ref{eq:NonEqId}) allow to obtain the adiabatic linear response relation (\ref{eq:NonEqIdLin5}), as well as the higher order terms in the adiabatic expansion. An interesting open question is whether and under which conditions geometric forces appear in those higher order terms.

Our main result, Eq. (\ref{eq:main}), provides a straightforward way to define
the Berry phase of an open quantum system. Just like the surface integral of the classical two-form (\ref{eq:clas2form}) provides a generalization of Berry phase for chaotic classical systems \cite{Robbins92PRSLA436,Jarzynski95PRL74}, so does the surface integral of the geometric magnetism  (\ref{eq:main}) provide an analogue of the Berry phase of open quantum systems,
reading
\begin{equation}
\gamma = \int \bm{\mathcal{B}} \cdot \mathrm{d}\bm\Sigma \, .
\end{equation}
This so defined phase $\gamma$ would in general differ from those, equally sound and useful, expressions of a Berry phase introduced for open systems in the prior literature  \cite{Carollo03PRL90,De-Chiara03PRL91,Tong04PRL93,Whitney05PRL94,Buric09PRA80,Bassi06PRA73,Dajka11QIP10,Marzlin04PRL93,Whitney03PRL90,Sinitsyn09JPA42,Ren10PRL104}. In full analogy with the original Berry phase, $\gamma$ is geometric, that is it depends only on the path described by the driving parameters. It vanishes for a path enclosing no area, and it vanishes in the case when the system dynamics are time-reversal invariant, i.e., when  for any $t$, $\Theta \mathcal{H}(\bm{R}_t)= \mathcal{H}(\bm{R}_t) \Theta$. Here $\Theta$ is the anti-unitary time-reversal operator which reverses momenta and keeps the spatial coordinates and all external parameters (possibly including physical magnetic fields) unaltered \cite{Messiah62Book,Campisi11RMP83}. This is so because, due to Onsager-Casimir relations \cite{Casimir45RMP17}, the conductance matrix $\bm{K}$ would be symmetric in this case, hence,
the geometric magnetism $\bm{\mathcal{B}}$ would vanish.

Our simple example of a quantum harmonic oscillator transported along a path already shows that the presence of an environment does not
destroy geometric magnetism. In fact, in this specific (linear) case the geometric magnetism is given by the actual physical magnetic field,
exactly like in the isolated case \cite{Robbins94JPA27,Exner00PLA276}.
For nonlinear systems the difference between the real  and geometric magnetic
fields could be detected, as well as the difference between quantum and classical regimes.
However, the quantum-mechanical treatment of nonlinear open systems constitutes an ambitious challenge
because in this case the system evolution
cannot be handled analytically in an exact manner. This challenge, in principle, could be approached numerically, for example, (i) by resorting
to the Floquet-Markov formalism \cite{Grifoni98PR304}, under the assumption of weak system-bath coupling;
or (ii)  by following the Feynmann-Vernon path integral formalism \cite{Feynman63AP24,Grabert88PREP168},
to calculate the reduced density matrix numerically, through stochastic unraveling of the corresponding influence functional \cite{Koch08PRL100}.

Geometric magnetism is at the basis of a currently growing experimental activity aimed at producing
\emph{artificial gauge fields} in thermally isolated cold atomic gases \cite{Bloch08RMP80,Lin09NAT462,Dalibard11RMP83,Struck12PRL108,Jimenez-Garcia12PRL108}. The present theory opens the possibility of engineering synthetic gauge fields also in the presence
of a thermal environment, via our general expression (\ref{eq:main}).

\section*{Acknowledgments} The authors wish to thank Peter Talkner for
helpful discussions about open quantum systems. This work was supported by the cluster of excellence Nanosystems Initiative Munich (NIM) and the Volkswagen Foundation (Project No. I/83902).

\end{document}